\documentstyle[11pt]{article}
\def\bea{\begin{eqnarray}}
\def\eea{\end{eqnarray}}

\def\beq{\begin{equation}}
\def\eeq{\end{equation}}
\def\ba{\beq\new\begin{array}{c}}
\def\ea{\end{array}\eeq}

\makeatletter
\newdimen\normalarrayskip              
\newdimen\minarrayskip                 
\normalarrayskip\baselineskip
\minarrayskip\jot
\newif\ifold             \oldtrue            \def\new{\oldfalse}
\def\arraymode{\ifold\relax\else\displaystyle\fi} 
\def\eqnumphantom{\phantom{(\theequation)}}     
\def\@arrayskip{\ifold\baselineskip\z@\lineskip\z@
     \else
     \baselineskip\minarrayskip\lineskip2\minarrayskip\fi}
\def\@arrayclassz{\ifcase \@lastchclass \@acolampacol \or
\@ampacol \or \or \or \@addamp \or
   \@acolampacol \or \@firstampfalse \@acol \fi
\edef\@preamble{\@preamble
  \ifcase \@chnum
     \hfil$\relax\arraymode\@sharp$\hfil
     \or $\relax\arraymode\@sharp$\hfil
     \or \hfil$\relax\arraymode\@sharp$\fi}}
\def\@array[#1]#2{\setbox\@arstrutbox=\hbox{\vrule
     height\arraystretch \ht\strutbox
     depth\arraystretch \dp\strutbox
     width\z@}\@mkpream{#2}\edef\@preamble{\halign
\noexpand\@halignto
\bgroup \tabskip\z@ \@arstrut \@preamble \tabskip\z@ \cr}%
\let\@startpbox\@@startpbox \let\@endpbox\@@endpbox
  \if #1t\vtop \else \if#1b\vbox \else \vcenter \fi\fi
  \bgroup \let\par\relax
  \let\@sharp##\let\protect\relax
  \@arrayskip\@preamble}
\def\eqnarray{\stepcounter{equation}%
              \let\@currentlabel=\theequation
              \global\@eqnswtrue
              \global\@eqcnt\z@
              \tabskip\@centering
              \let\\=\@eqncr
              $$%
 \halign to \displaywidth\bgroup
    \eqnumphantom\@eqnsel\hskip\@centering
    $\displaystyle \tabskip\z@ {##}$%
    \global\@eqcnt\@ne \hskip 2\arraycolsep
         $\displaystyle\arraymode{##}$\hfil
    \global\@eqcnt\tw@ \hskip 2\arraycolsep
         $\displaystyle\tabskip\z@{##}$\hfil
         \tabskip\@centering
    &{##}\tabskip\z@\cr}

\begin{document}

\begin{titlepage}
\setcounter{footnote}0
\begin{center}
\hfill ITEP/TH-18/98\\

\vspace{1.3in}
{\LARGE\bf On a Casher-Banks relation in MQCD}
\date{today}

\bigskip {\Large A.Gorsky}

\bigskip { ITEP, Moscow, 117259, B.Cheryomushkinskaya 25}
\\
\end{center}
\bigskip

\begin{abstract}
We discuss the meaning of a Casher-Banks
relation for the Dirac operator eigenvalues in MQCD. It suggests
the interpretaion of the eigenvalue as a coordinate involved
in the brane configuration.
\end{abstract}

\end{titlepage}


1. Brane approach to SUSY gauge theories in the different dimensions
seems to be the most promising tool for capturing nonperturbative features
at the strong coupling regime. It is believed that all
universal characteristics of the vacuum sector can be seen or calculated
in terms of brane configuration. Recently proper brane configuration was
found in IIA \cite{kutasov}  and M
theory \cite{ooguri,witten1} for the theories
with N=1 supersymmetry. Configuration for the pure gauge theory
suggested in \cite{kutasov} involves NS5 and D4 branes which being lifted
to M theory are identified with the single M5 brane wrapped around Riemann
surface embedded into three dimensional complex space.

It is known that gluino condensate is developed  and properties
of the vacuum sector in N=1 SQCD are governed by superpotentials generated
in different ways depending on the relation between $N_{f}$ and $N_{c}$.
Gluino condensate indicates the chiral symmetry breaking so
clarification of its origin
in MQCD framework is very important. It was found that proper
superpotential can be calculated in M theory indeed \cite{witten1,ooguri,nam}
moreover it turns out that superpotential can be attributed to the 5 brane
instantons \cite{wittensup,gomez}. General consideration of the chirality
along the brane approach indicates that at least in some situations it
depends only on the local properties of the brane configuration
\cite{hb}. One more important observation was made in \cite{shifman}
that algebra of N=1 SUSY implies the existence of the domain wall
between the space regions with different phases of the gluino condensate.

In this note we consider the Casher-Banks relation \cite{CB}
relating the properties of the Dirac operator eigenvalues in the
instanton ensemble background to the value of the vacuum fermion
condensate. This relation has been known in QCD for a while where
fermions in the fundamental representation condense
in the vacuum. Generalization to the adjoint fermions relevant for
the supersymmetric theory as well as additional sum rules for the
spectral density of the Dirac operator have been discovered
in  \cite {ls}. Spectral density is the universal characteristics
of the low energy sector of the theory so one expects that
it can be also treated in the brane terms. Note that universal behaviour
of the spectral density admits  matrix model approach for the
investigation of its properties (see \cite{ver} for a review )
which sucsessfully describes key features of the spectrum.
Below we suggest interpretation of a Casher-Banks relation
in terms of M5 brane worldvolume and discuss the possible
role of the Dirac operator eigenvalue as one of the dimensions
involved in the brane picture.

2. Let us remind the main facts about the spectrum of  Dirac
operator in QCD. One concerns eigenvalues of the
operator $\hat{D}\Psi=\lambda\Psi$ and density of the eigenvalues
$\rho(\lambda)$ provides the order parameter in the low energy sector.
Namely due to Casher-Banks relation \cite{CB} for the fundamental fermions
\beq
<\bar {\Psi}\Psi>=\pi \rho(0)=\int \frac{\rho(\lambda)d\lambda}{\lambda} ,
\eeq
density at origin can be considered as the derivative of the partition
function with respect to the mass
$\pi \rho(0)=\frac{dlnZ}{dm}$ at $m=0$. More generally, there is an infinite
tower of  sum rules for the inverse powers of eigenvalues \cite{ls},
for instance

\beq
<\sum \frac{1}{\lambda_{i}^{2}}>_{\nu}=
\frac{(<\bar {\Psi}\Psi> V)^{2}}{4(\mid \nu \mid + N_{f})},
\eeq
where the averaging with the spectral density  is implied,
V is the four dimensional Euclidean volume and $\nu$ is the
topological charge of the gauge field configuration.
Sum rules probe the
structure of the small $\lambda$ region in the spectrum.

One more approach to the same quantity comes from the representation
of the partition function as the path integral averaged fermion
determinant
\beq
Z(m)=<Det(iD-m)>.
\eeq
Averaging can be substituted by the integration
over the instanton moduli space which due
to ADHM description is modelled by the proper matrix model \cite{ver}.
Along this way of reasoning it is possible to derive
the eigenvalue distribution itself.
Actually determinant can be considered as
the observable in the theory dealing with instanton moduli space.
It is generally believed that in the QCD instanton vacuum spectrum
asquires the band structure due to the delocalization of the zero
mode on the single instanton in the instanton ensemble. It is assumed
that zero eigenvalue lies in the allowed band providing
the chiral symmetry breaking.

A little bit different way to capture the fermion condensate is
to consider the analytic properties of the resolvent of the Dirac
operator considered on the complex mass plane
\beq
G(z)=<Tr\frac{1}{iD-z}> .
\eeq
Riemann surface of G(z) has the cut
along the imaginary axis which is in one-to-one correspondence with
the presence of the condensate.

3. Let us now proceed to the supersymmetric case. To start note
that now we consider the condensate of gluinos - fermions
in the adjoint representation.
Casher-Banks relation for the
adjoint fermions holds true  \cite{ls}
so the issue of the derivation of the gluino condensate via
the spectral density is well defined.

We would like to get the brane interpretaion of the
Casher-Banks relation as well as to obtain some features
of the spectral density. We will use the M-theory picture
to get the SUSY gauge theory on the world-wolume of the
M-theory fivebrane whose 2 dimensional part in N=2 case is wrapped
around the surface $\Sigma$   \cite{wit2} (see also \cite{vafa})
\beq
t^{2}-P_{n}(v)t+1=0
\eeq
In what follows we would like to present the arguments that $v$ can
be identified with the eigenvalue of the 4d Dirac operator.
To get N=1 theory we should proceed in two steps.
First, we have to shrink all monopole cycles on the surface $\Sigma$
reducing the curve  to
\ba
y^2=(\frac{v^{2}}{4}-1)Q^{2}_{n-1}(v)  \\
y=t-t^{-1},
\ea
where $Q_{n}$ - Chebyshev polynomials. Then one performs "rotation"
which embedds the new curve into three dimensional complex space.
The resulting rational curve has the form
\beq
v=t^{n}; vw=\theta ,
\eeq
where $\theta$ was identified with the gluino
condensate in \cite{ooguri,witten1}. We would like
to consider the equation
\beq
<\xi\xi>=\int wdv ,
\eeq
where $\xi$ is the gluino field,
as Casher-Banks relation.
Since in the IIA language  D4 branes are located between two NS5 branes
this identification
implies that the condensate develops purely due to the presense of one  NS5
brane at $v=0$.

Let us comment on the interpretation of $v$,$w$ and $t$ variables in the
field theory framework. In the theory with N=4  supersymmetry it
is straightforward to identify six dimensions with zero modes of
three complex scalar fields. If one treats N=2  theories
starting from softly broken N=4 ones then  trace of
two complex dimensions
corresponding to the massive scalar fields survives due to the dimensional
transmutation. We conjectured quite different interpretation
in N=1 theory assuming that $v$ is the eigenvalue of the Dirac
operator. Therefore
coordinates $w$ and $t$ have to be considered as the eigenvalues of
other operators commuting with the Dirac one and
can be associated with the global symmetries.

Let us compare our interpretation of N=1 theory with N=2 one
and  show that there is qualitative agreement with the standard
treatment of these theories. We will use arguments coming from
approach based on the relation of N=2 theories to the
integrable systems \cite{gkmmm}.

In IIA picture N=2 theory lives on the worldvolume
of $N_{c}$ D4 branes with finite extent in $x_{6}$ dimension \cite{wit2}.
Parallel NS5 branes are located at $x_{6}=0$ and
$x_{6}=\frac{l_{s}}{g^{2}g_{s}}$ where $g_{s}$ and $l_{s}$ are IIA string
coupling and length respectively.
It was argued in \cite{g,ggm} that integrable structure behind the
solution of the theory implies that
there are $N_{c}$ D0 branes living on D4 one per each
whose equilibrium positions are
\beq
x_{6,k}=\frac{kl_{s}}{N_{c}g^{2}}.
\eeq
Being lifted to the M theory D0 branes represent the linear bundle
on the spectral curve of the integrable system, which has been
identified with the Riemann surface  $\Sigma$ which M5 brane is wrapped
around.  Linear bundle yields half of the phase space of the integrable
system and fluctuations of interacting D0 branes around these points
define just
Toda dynamics which linearizes on the Jacobian of the spectral curve. These
branes can be  treated as the point-like abelian instantons in 4d theory and
their possible role was discussed in \cite{bp,brodie,lust}.

Let us consider now fermions on M5 worldvolume. We will
look for the zero mode of 6d Dirac operator D which we decompose as
\beq
D=D^{4}+d^{\Sigma},
\eeq
where $d^{\Sigma}$ denotes the Dirac operator on the spectral curve.
We decompose fermion wave function  after IIA projection as
\beq
\Psi(x,x_{6})=\sum_{\lambda} \Psi_{\lambda}(x)\Phi(n,\lambda) ,
\eeq
where $x=(x_{0},x_{1},x_{2},x_{3})$ and n denotes the sites where
D0,s are localized.
We conjecture that $\Phi(n,\lambda)$ is just the Baker function
for  the Lax equation in Toda system if the
eigenvalue of the four dimensional Dirac operator $D^{4}$ equals $\lambda$.
Baker fermions in the Toda system  have $\Sigma$ as the Fermi surface.

In IIA projection one has the fermion zero modes localized on $D0^{,}s$
so it is natural to consider fermion wave function $\Phi(n,\lambda)$
with discrete n. Fermion zero modes on the nearest
$D4^{,}s$ overlap and the Dirac operator
along $x_{6}$ acquires the discrete form
\ba
c_{n}\Phi_{n-1}+p_{n}\Phi_{n}+c_{n+1}\Phi_{n+1}=\lambda\Phi_{n}  \\
c_{n}=exp(x_{n+1}-x_{n}) ,
\ea
where $x_{i}$ is the position of i-th D0 brane along $x_{6}$ direction,
and coincides with the familiar expression for the Lax equation
in the Toda system in $2\times2$ representation.

Let us emphasize that Hitchin like dynamical system involving
D0 branes can not be seen at the classical level and has to be regarded as
the quantum or quasiclassical effect. Phase space of
the relevant Hitchin system is the hyperkahler manifold and can be
interpreted as a hidden Higgs branch of the moduli space in N=2 theories.
The very meaning of integrability is to restrict
$N_{c}$ KK modes in M theory
picture to the M5 brane worldvolume.

Comparing expressions for the condensate in N=1 theory we expect the
identification
\beq
w(v)dv \propto \frac{\rho(\lambda)}{\lambda}d\lambda.
\eeq
It is worth to comment on the corresponding deformation of the
integrable system. At the first step we should fix all integrals
of motion in the Toda system and then perturb it by the mass term.
It is not clear at the moment if some integrable dynamics survives
but the potential degrees of freedom namely $D0^{,}s$ still play
important role in generation of the superpotential \cite{brodie}.
Therefore one  has at least proper candidates for the phase space -
rational spectral curve which M5 brane is wrapped
around and the linear bundle on it.

4. Some insights come from the analogy with the
discrete Peierls model \cite{gpei}, which
is relevant for description  of 1d superconductivity. The model is defined
as follows; there is a periodic 1d crystal with $N_{c}$ sites and fermions
interacting with phonon degrees of freedom. It is assumed that
fermions are strongly coupled to the lattice and can jump only
between the nearest sites. This system is described by the periodic
Toda lattice \cite{bdk}, where the Toda Lax operator serves as the
Hamiltonian for fermions and Toda degrees of freedom amounts from
the phonons.

Spectral curve for the Toda system coincides with the dispersion law of the
fermions and Lax equation - with the Shr$\ddot o$dinger one.
Initial condition for the Toda evolution is chosen
dynamically to minimize the total energy of fermions and  phonons
and it appears that the generic curve degenerates to the "N=2 curve"
with all monopole cycles vanish just as in the first step towards
N=1 theory. Fermions develop the mass gap  moreover the analogue
of the chiral invariance known in the model is broken dynamically
due to the survived band around  zero energy. This yields a dynamical
scenario which resembles one in the N=1 theory.

To conclude, we have considered the
possible meaning of the Casher-Banks relation
in MQCD framework. We obtain qualitative picture explaining
the  nonvanishing spectral density of the Dirac operator at the
origin and discuss the geometrical meaning of the eigenvalues
themselves. We conjecture that six dimensions involved in the
brane configuration for N=1 theory
allow interpretation as the Dirac operator
eigenvalue and eigenvalues of the two global
symmetry generators.
Certainly more quantitative analysis is required to
confirm the picture suggested.

We would like to thank H.Leutwyler, A.Mironov, A.Morozov and A.Smilga for
the useful discussions. This work was supported in part by
grants CRDF-RP2-132, INTAS 96-482 and RFFR-97-02-16131.

\end{document}